\documentclass[sigconf]{acmart}

\usepackage{booktabs} 
\usepackage{hhline}
\usepackage{array}
\usepackage{afterpage}
\usepackage{placeins}
\usepackage{color}
\usepackage{soul}
\usepackage{graphicx}
\usepackage{lipsum}
\usepackage{subfig}
\usepackage{caption}
\usepackage{tabularx}
\usepackage[nolist,nohyperlinks,smaller]{acronym}
\usepackage{tikz}
\usepackage{balance}
\usepackage{svg}

\renewenvironment{quote}{%
\list{}{%
\leftmargin0.4cm 
\rightmargin\leftmargin
\topsep4pt
}
\item\relax
}
{\endlist}

\begin{acronym}
    \acro{AR}{Augmented Reality}
    \acro{VR}{Virtual Reality}
    \acro{3D}{three-dimensional}
    \acro{2D}{two-dimensional}
    \acro{VCS}{Version Control System}
    \acro{DAG}{Directed Acyclic Graph}
    \acro{HG}{History Graph}
    \acro{GUI}{Graphical User Interface}
    \acro{WIM}{World in Miniature}
    \acro{LLM}{Large Language Model}
\end{acronym}

\newcommand{\sys}{VRCopilot}

\AtBeginDocument{%
  \providecommand\BibTeX{{%
    \normalfont B\kern-0.5em{\scshape i\kern-0.25em b}\kern-0.8em\TeX}}}

\copyrightyear{2024}
\acmYear{2024}
\setcopyright{acmlicensed}
\acmConference[UIST '24]{The 37th Annual ACM Symposium on User Interface Software and Technology}{October 13--16, 2024}{Pittsburgh, PA, USA}
\acmBooktitle{The 37th Annual ACM Symposium on User Interface Software and Technology (UIST '24), October 13--16, 2024, Pittsburgh, PA, USA}
\acmDOI{10.1145/3654777.3676451}
\acmISBN{979-8-4007-0628-8/24/10}

\begin{document}
\title[VRCopilot: Authoring 3D Layouts with Generative AI Models in VR]{VRCopilot: Authoring 3D Layouts with \\Generative AI Models in VR}


\author{Lei Zhang}
\affiliation{%
 \institution{University of Michigan}
 \city{Ann Arbor}
 \state{MI}
 \country{USA}}
 \email{raynez@umich.edu}

\author{Jin Pan}
\affiliation{%
 \institution{University of Michigan}
 \city{Ann Arbor}
 \state{MI}
 \country{USA}}
 \email{jhinpan@umich.edu}

\author{Jacob Gettig}
\affiliation{%
 \institution{University of Michigan}
 \city{Ann Arbor}
 \state{MI}
 \country{USA}}
 \email{jgettig@umich.edu}

\author{Steve Oney}
\affiliation{%
 \institution{University of Michigan}
 \city{Ann Arbor}
 \state{MI}
 \country{USA}}
 \email{soney@umich.edu}

\author{Anhong Guo}
\affiliation{%
 \institution{University of Michigan}
 \city{Ann Arbor}
 \state{MI}
 \country{USA}}
 \email{anhong@umich.edu}

\newcommand{\lei}[1]{\textcolor{black}{#1}}

\begin{abstract}

Immersive authoring provides an intuitive medium for users to create 3D scenes via direct manipulation in Virtual Reality (VR). Recent advances in generative AI have enabled the automatic creation of realistic 3D layouts. However, it is unclear how capabilities of generative AI can be used in immersive authoring to support fluid interactions, user agency, and creativity. We introduce VRCopilot, a mixed-initiative system that integrates pre-trained generative AI models into immersive authoring to facilitate human-AI co-creation in VR. VRCopilot presents multimodal interactions to support rapid prototyping and iterations with AI, and intermediate representations such as wireframes to augment user controllability over the created content. Through a series of user studies, we evaluated the potential and challenges in manual, scaffolded, and automatic creation in immersive authoring. We found that scaffolded creation using wireframes enhanced the user agency compared to automatic creation. We also found that manual creation via multimodal specification offers the highest sense of creativity and agency.
\end{abstract}



\begin{teaserfigure}
  \includegraphics[width=\textwidth]{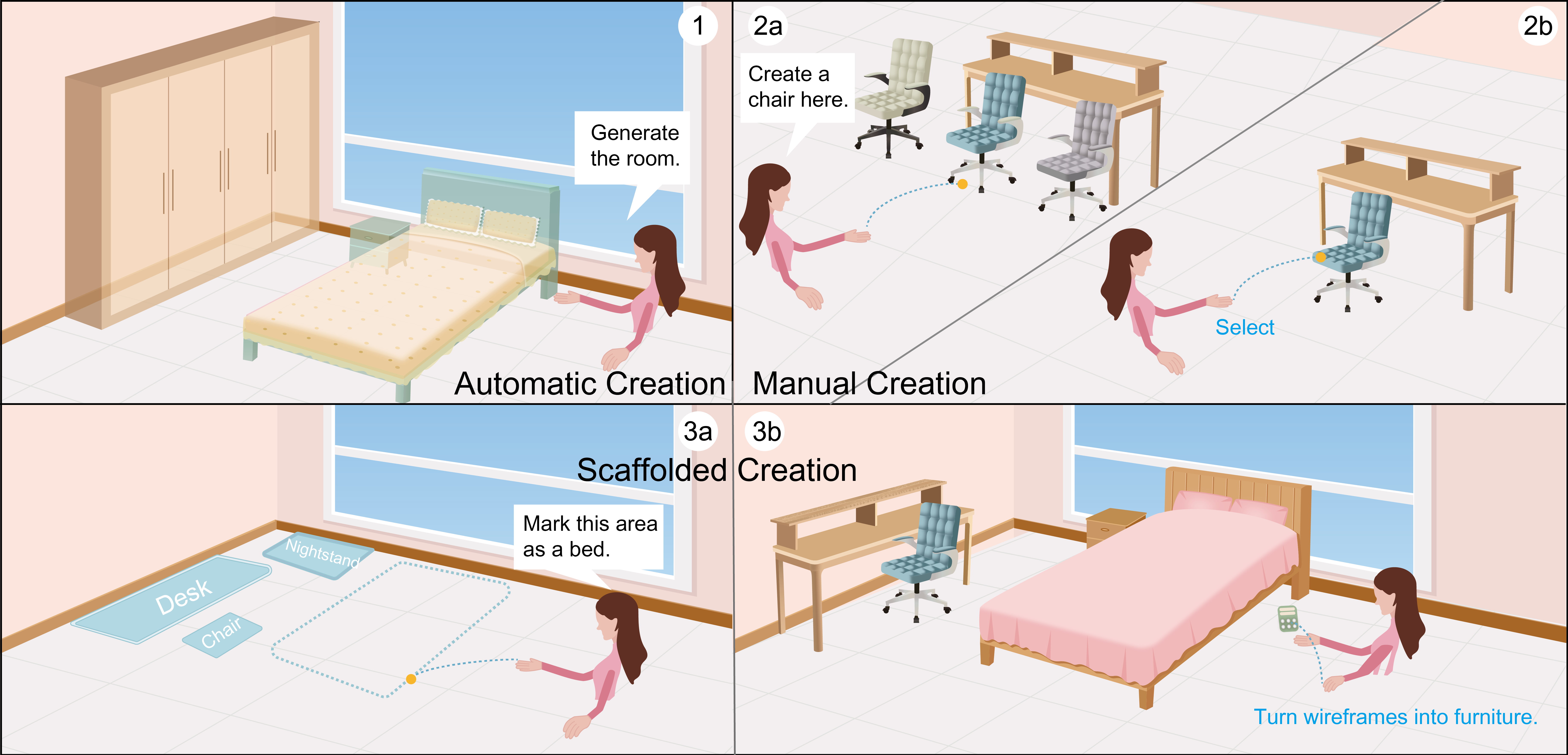}
  \caption{System Overview of VRCopilot. 1) Automatic Creation: Users can use voice commands to ask the generative model to generate a full-room layout based on an empty room. 2) \lei{Manual} Creation: Users can use multimodal specification by speaking with simultaneous pointing to ask the system to \lei{suggest} a chair (a); they can select from one of the three suggestions offered by the system (b). 3) Scaffolded Creation: Users can create \textit{wireframes} by drawing on the floor while speaking, in addition to automatically generated wireframes (a); They can then turn the wireframes into specific furniture (b).}
  \Description[System overview of VRCopilot]{Figure 1 shows the system overview of VRCopilot. 1) Automatic Creation: Users can use voice commands to ask the generative model to generate a full-room layout based on an empty room. 2) Manual Creation: Users can use multimodal specification by speaking with simultaneous pointing to ask the system to suggest a chair (a); they can select from one of the three suggestions offered by the system (b). 3) Scaffolded Creation: Users can create wireframes by drawing on the floor while speaking, in addition to automatically generated wireframes (a); They can then turn the wireframes into specific furniture (b).}
  \label{fig:teaser}
\end{teaserfigure}

\maketitle

\section{Introduction}

As \ac{VR} continues to gain momentum across various domains, such as education~\cite{zhu2023learniotvr}, gaming~\cite{unreal}, and spatial design~\cite{zhang2023vrgit}, the need for effective tools and techniques to author high-quality 3D scenes becomes increasingly important. 
Immersive authoring is a paradigm that leverages users' spatial capabilities to enable them to create and evaluate 3D scenes directly while immersed in the virtual environment~\cite{adobe, google, mine1995isaac, jackson2016lift, ens2017ivy, zhang2019studying, zhang2020flowmatic, zhang2023vrgit}.
Prior work in immersive authoring tools has demonstrated benefits of lowering the barrier for end-users with little technical skills to create 3D content~\cite{lee2004immersive, zhang2020flowmatic}.

While existing immersive authoring tools make it intuitive for users to visualize their design concepts for 3D scenes in VR, most current 3D layouts such as architectural designs and game scenes are laboriously created through manual placement of 3D models.
This manual process is not only tedious and time-consuming, but can also limit the user's ability to explore a diverse range of ideas~\cite{JANSSON19913}.
In recent years, generative Artificial Intelligence (AI) models have emerged as powerful means for automatic generation of intelligible text~\cite{gpt}, photorealistic images~\cite{pmlr-v48-reed16}, videos~\cite{sora}, music~\cite{louie2020novice}, and 3D layouts~\cite{paschalidou2021atiss, li2019grains, wang2019planit}.
By leveraging generative models, we can potentially provide users with automatically generated 3D layouts during the process of immersive content creation, enabling users to save time and effort while exploring alternative design possibilities. 

Prior work has demonstrated promising results in generating realistic 3D layouts~\cite{paschalidou2021atiss,li2019grains, wang2019planit} and text-to-layout generation~\cite{ma2018language, feng2024layoutgpt}.
However, integrating these models into immersive authoring workflows poses unique challenges of how users can collaborate and interact with generative models---specifically, understanding, controlling, and refining model outputs in immersive virtual environments.
This difficulty is compounded by generative AI models' well-known issues with transparency, controllability, and user agency.
Current generative models for 3D layouts use either room sizes (e.g., \cite{paschalidou2021atiss}) or text captions (e.g., \cite{feng2024layoutgpt}) as prompts.
It is difficult for users to define their design objectives, such as requesting layout designs with elements in particular locations or sizes (as seen in Fig. 1.3.b).

In this paper, we introduce VRCopilot, a mixed-initiative system that integrates pre-trained generative models into immersive authoring workflows.
VRCopilot is instantiated in the context of layout design for indoor scenes, where users are able to co-create with generative models via requesting, controlling, and refining generative models' outputs in VR.
VRCopilot introduces two key interaction techniques: \textit{(1) multimodal specification} and \textit{(2) intermediate representation}.
Inspired by multimodal interactions such as ``Put-that-there''~\cite{bolt1980put}, our system enables users to use speech and simultaneous pointing to specify their creation needs, increasing the naturalness and economy of language description in the immersive environments.
For instance, users can point to a location in the room while saying ``create a wooden chair here.''
As a response, the system will \lei{offer three options} for the user to choose from.
Besides, to help users co-create with the generative model in a more transparent and controllable way, VRCopilot proposes the notion of \textit{wireframes} as intermediate representations for the generated outcomes.
Inspired by the concept of low-fidelity prototyping in Human-Computer Interaction~\cite{rettig1994prototyping, buxton2010sketching, snyder2003paper}, wireframes are 2D representations of 3D layouts similar to floor plans in interior design.
These representations can be hand-drawn by users together with speech specifying the their types, or suggested by generative models.
VRCopilot allows users to iteratively refine the design with generative AI by enabling them to convert between intermediate representations and 3D layouts.

Taking the above techniques together, we propose three ways of human-AI co-creation in VR enabled by VRCopilot: (1) \lei{manual} creation, where users create individual objects to complete a layout design via a catalog menu and multimodal specification; (2) automatic creation, where users request and refine suggestions from generative AI for full-room layouts; and (3) scaffolded creation, where users co-create intermediate representations with generative AI for guiding the final layout design.

To provide an in-depth understanding of the human-AI co-creation process in VR, we conducted two rounds of user studies.
Our first study aimed to compare user experiences of creating 3D layouts with and without AI.
Specifically, we compared creation without AI using \lei{manual} placement and creation with AI using generative models.
We found that co-creating 3D layouts with generative models is generally more preferable as it could save users' effort while resulting in 3D layouts with more complete \textit{functionality} and diverse \textit{color palette}.
However, users struggled with the generative model's non-deterministic output, where the generated results might misalign with the user's design goals due to the lack of controllability of the generative model.

Based on the insights and challenges from the first study, we further evaluated VRCopilot by comparing different levels of AI automation in the creation process including \lei{manual} creation, automatic creation, and scaffolded creation.
\lei{We found that users' sense of agency significantly increases in the order of automatic creation, scaffolded creation, and manual creation.
Specifically, the design of wireframes in scaffolded creation enhances users' agency by allowing them to define the 3D layout including object types and sizes compared to automatic creation.
Manual creation offers the highest agency by enabling additional visualization and control over object styles.
We also found that users felt significantly more creative in manual creation, than in scaffolded or automatic creation, with no significant difference found between the latter two.
Specifically, having multiple suggestions via multimodal specification in manual creation can make users feel more creative.
Users felt less creative in the other two conditions since AI generated outcomes could lead to fixation and prohibit users from creative exploration.}

In sum, our paper makes the following contributions: 1) VRCopilot, an immersive authoring system that enables users to interact and co-create with generative AI models in virtual immersive environments; and 2) empirical results gained from two user studies that provide insights on user experiences such as perceived agency and creativity, as well as potential and challenges of human-AI co-creation in immersive authoring workflows.

\section{Related Work}

\sys{} draws inspiration from prior literature on 3D scene synthesis using generative models, creativity support with generative design, and interactive interfaces with computational agents.

\subsection{Generative Models for 3D Scenes}
The demand for automatically generating 3D scenes has never been higher in the domain of gaming, AR \& VR, architecture and interior design.
\lei{In the field of computer vision, this topic named 3D scene synthesis is gaining popularity and prior researchers have explored generating new 3D scenes via various input including images~\cite{liu2023zero, gao2022get3d}, text~\cite{zeng2022lion, deitke2024objaverse}, or room shape~\cite{paschalidou2021atiss}.
A key line of work is 3D \textit{indoor} scene synthesis, which refers to the task of automatically generating a set of 3D furniture objects along with their positions and orientations, given a room layout~\cite{zhang2019survey}.}
Some of the early work in this space offered suggestions using hardware-accelerated Monte Carlo sampler based on interior design guidelines~\cite{merrell2011interactive}.
Follow up work has been focused on data-driven approaches, given the rise of large 3D object datasets such as SUNCG~\cite{song2017semantic} and 3D-FRONT~\cite{fu20213d}.
The data-driven approaches can be approximately categorized into graph-based~\cite{wang2019planit} and autoregression-based approaches~\cite{wang2018deep, ritchie2019fast, wang2021sceneformer, paschalidou2021atiss}.
Graph-based approaches encode 3D layouts as scene graphs, where objects are nodes, and the spatial relationship between objects are edges.
This method treats the task of generating 3D scenes as generating directional graphs.
The main motivation behind this is to process it with graph convolutional networks.
Most notably, Ritchie et al.~\cite{ritchie2019fast} introduced a CNN-based architecture that operates on a top-down image-based representation of a scene and inserts objects in it sequentially by predicting their category, location, orientation, and size.
More recently, autoregression-based approaches have been introduced.
Wang et al. introduced SceneFormer~\cite{wang2021sceneformer}, a series of transformers that autoregressively add objects in a scene.
ATISS~\cite{paschalidou2021atiss} simplifies the process by proposing a single model trained end-to-end to predict all attributes.
Most notably, ATISS encodes 3D objects' positions, rotations, and scales in transformers for training.
\lei{More recently, DiffuScene utilizes a denoising diffusion model that is able to generate more plausible and diverse indoor scenes~\cite{tang2024diffuscene}.}

Our work contributes to the existing literature on 3D scene synthesis by introducing generative models into immersive environments.
While prior work has been focused on generating realistic 3D layouts, VRCopilot aims to integrate state-of-the-art generative AI models into immersive authoring and explores the ways of co-creating with generative AI models in VR.

\subsection{Creativity Support via Steering Generative Models}
The acceleration of AI capabilities has enabled human-AI co-creation in domains such as drawing~\cite{davis2016empirically, fan2019collabdraw}, creative writing~\cite{clark2018creative}, video game content creation~\cite{guzdial2019friend}, and music composition~\cite{huang2019bach, louie2020novice}.
For example, Bach Doodle~\cite{huang2019bach} is able to complete a music composition in the style of J.S. Bach by requiring users to only write a few notes.
While recent research has focused on building co-creation experiences in 2D interfaces, there has been relatively little HCI work examining how to design interactions with these state-of-the-art generative models to ensure they are effective for co-creation in the immersive environments.
Our research contributes an understanding of how interactions with these AI models can be designed, how they affect the immersive authoring experience, and users' attitudes towards AI co-creation in VR.

Integrating existing generative AI models into creative work presents unique challenges in itself such as adapting actions of AI based on users' preferences~\cite{chung2022talebrush, kazi2017dreamsketch, swearngin2020scout}.
Research has also observed that users desire to take initiative in their partnership with AI, and thus sought to provide steering tools to make AI align with users' creative goals.
For example, TaleBrush~\cite{chung2022talebrush} uses a combination of line sketching and natural language narration to create stories.
DreamSketch~\cite{kazi2017dreamsketch} uses sketches as input for the generative design of 3D models.
In the domain of 2D layouts, Scout~\cite{swearngin2020scout} uses high-level constraints based on design concepts to generate multiple designs.
Building on this need, our work investigates how users express their preferences to generative AI through multimodal specification and intermediate representations in VR.

\subsection{Interaction Techniques in Immersive Environments}
Our proposed interactions are inspired by prior interaction techniques in immersive environments including multimodal interaction, spatial interaction, and world in miniature (WIM).
While recently there has been extensive exploration in using natural language interactions with generative AI models to build virtual scenes, using just natural languages alone might be effective for tasks such as referencing~\cite{chastine2007understanding} in immersive environments.
Building on this line of work, our system demonstrates how multimodal interaction can be used for specifying objects to generative AI in the immersive authoring process.
Finally, Stoakley et al. introduced the concept of World in Miniature (WIM), which enables both
navigation and interaction in a large VR scene~\cite{stoakley1995virtual}. 
A WIM represents the virtual environment and allows users to manipulate objects offered by the miniature, or rapidly teleport in the virtual environment by selecting locations directly in the miniature. 
It also has the benefit of allowing users to see a preview of the immersive virtual environment without having to travel back and forth between different views.
We built on the WIM technique to enalbe users to design and edit multiple variations of the 3D layouts.

\section{VRCopilot}

\sys{} is a mixed-initiative immersive authoring system that enables users to co-create 3D layouts with pre-trained generative models in VR.
Users can ask generative models to generate full-room layouts or \lei{use multimodal specification to create} individual objects.
They can also manually place objects from a catalog menu or request suggestions from the system using multimodal interactions (i.e., pointing and speaking).
\sys{} further allows users to create \emph{wireframes} --- intermediate representations that help guide and refine the layout generation process.
We detail our system design in the sections below.

\subsection{Scope}
We situate our design of \sys{} in the context of interior design tasks, where users can place pre-made 3D furniture models in a virtual apartment.
Interior design requires balancing constraints (e.g., functional requirements and space limitations) with aesthetic preferences.
It has been the application domain of many prior immersive authoring tools~\cite{ibayashi2015dollhouse,chow2019challenges,zhang2023vrgit} and is a common use case for Mixed Reality.
For example, several popular home goods stores, including IKEA\footnote{\url{https://www.ikea.com/global/en/newsroom/innovation/ikea-launches-ikea-place-a-new-app-that-allows-people-to-virtually-place-furniture-in-their-home-170912/}}, integrate features that allow customers to virtually preview furniture arrangements in their own homes before making a purchase. 
\sys{} includes $7,302$ furniture models from 3D-FRONT~\cite{fu20213d}, a large open-source dataset of furniture objects and textures.

Designing \sys{} for interior design allows us to evaluate it in a realistic domain with demonstrated utility.
However, we believe many of the concepts behind our design could generalize to other spatial design tasks, as the low-level tasks (e.g., object instantiation, customization, and manipulation) and multimodal interactions with generative models are broadly applicable across domains.

\begin{figure}[t!]
  \centering
  \vspace{-.5pc}
  \subfloat[Palette Menu.\label{fig:authoring_interface_1}]{\includegraphics[clip,width=0.368\linewidth]{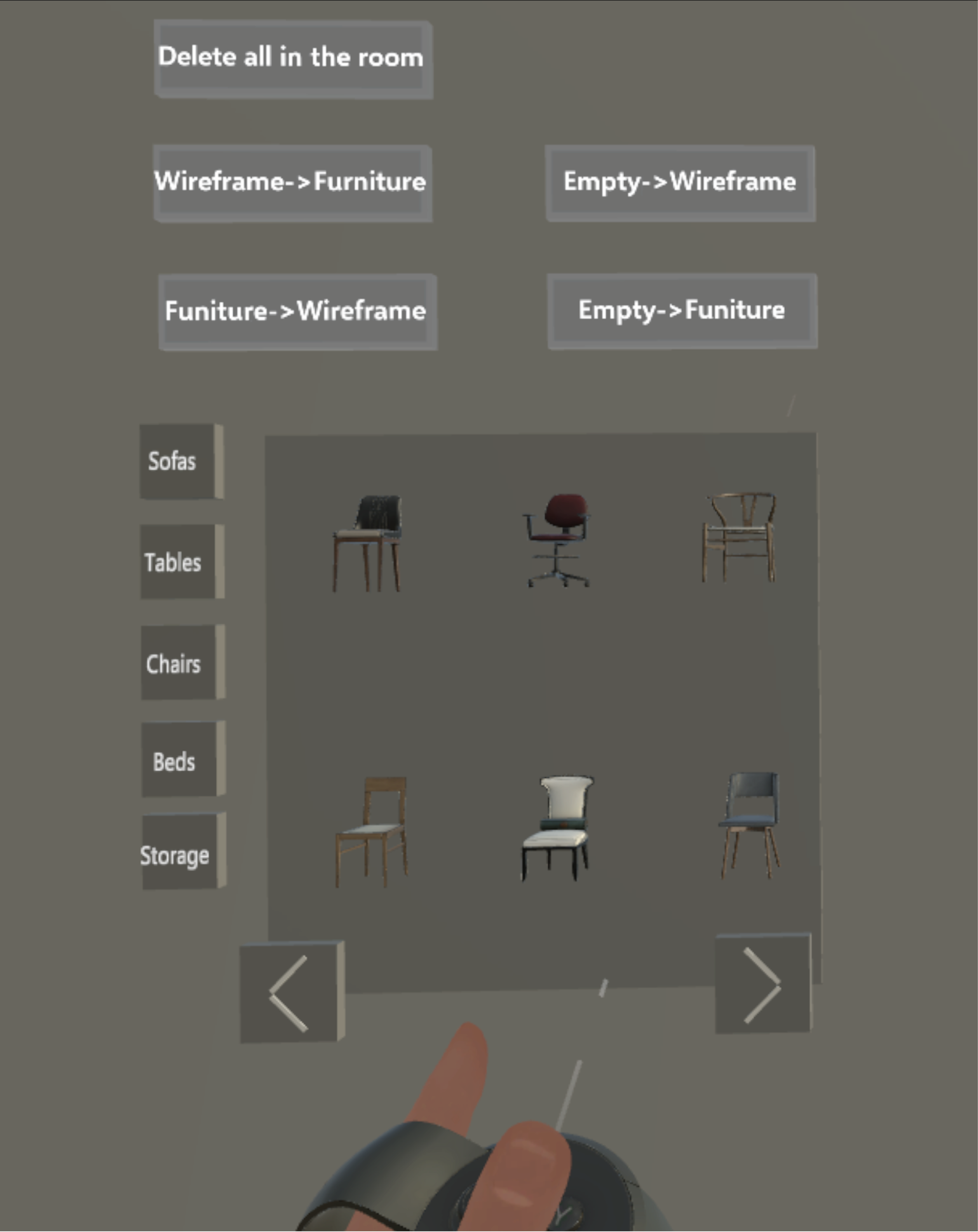}} 
  \subfloat[Multi-workspace.\label{fig:authoring_interface_2}]{\includegraphics[clip,width=0.632\linewidth]{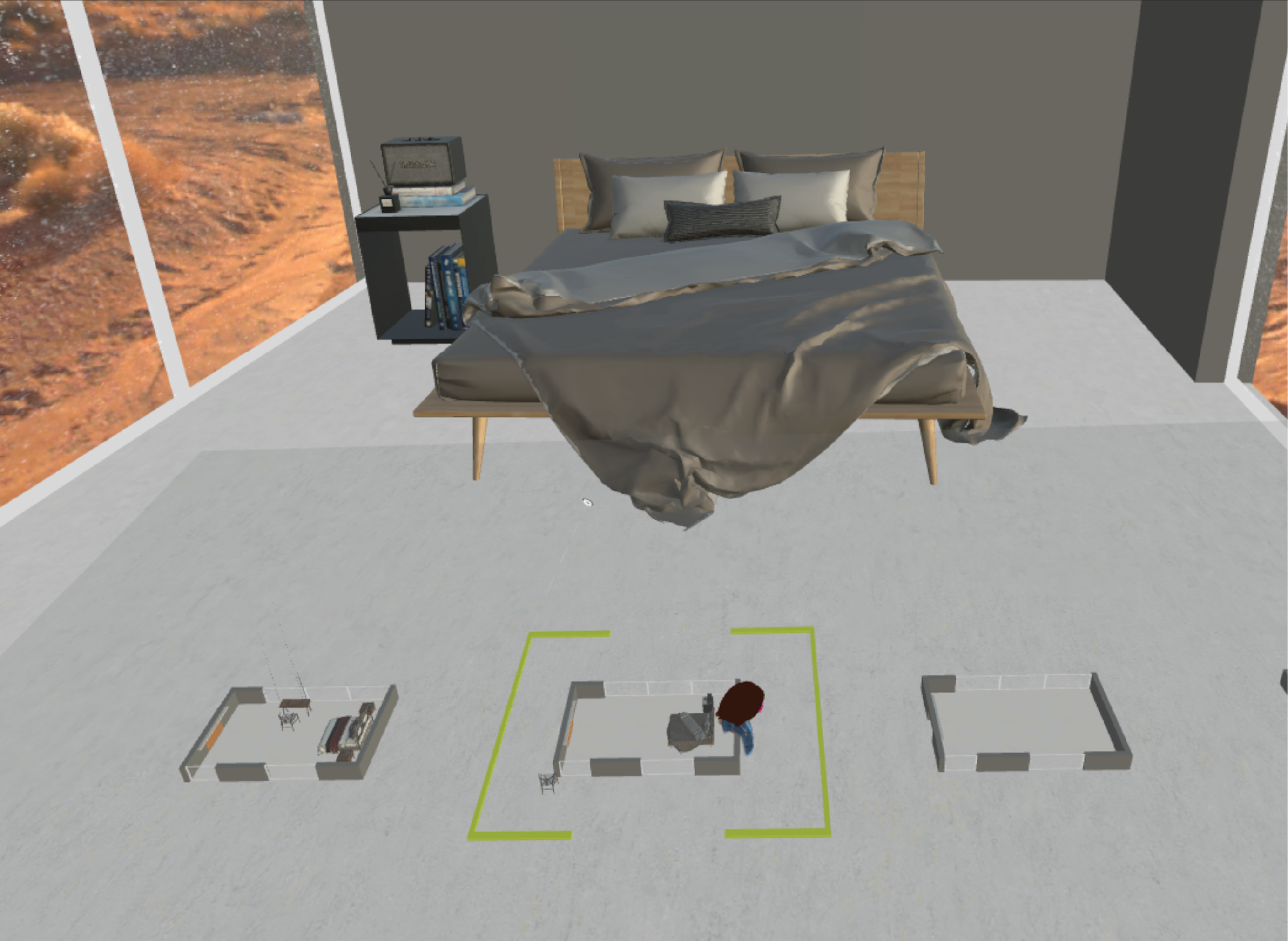}}
  \vspace{-.5pc}
  \caption{User interfaces in VRCopilot including (a) a palette menu where users can select and furniture from the catalog, and (b) a multi-workspace visualization that allows users to work and switch between multiple versions.}
  \Description[User interfaces in VRCopilot]{Figure 2 shows the user interfaces in VRCopilot including (a) a palette menu where users can select and furniture from the catalog, and (b) a multi-workspace visualization that allows users to work and switch between multiple versions.}
  \label{fig:authoring_interface}
  \vspace{-.5pc}
\end{figure}

\subsection{Immersive Authoring Features}
\sys{} is designed as an \emph{immersive authoring} tool, meaning users design a room layout while immersed in that room (in VR).

\subsubsection{Importing Models}
Users can manually import furniture models into the virtual environment from a catalog menu bound to their non-dominant hand, as seen in Fig. \ref{fig:authoring_interface_1}.
Using the catalog menu, users are able to choose from different categories of furniture such as tables and chairs from a sub-menu.
Each page on the catalog menu contains \lei{six} furniture items and they can turn pages to navigate more items.
After a furniture model is imported, users can manipulate and place the model via direct manipulation using the VR controllers.

\subsubsection{Design Exploration}
The ability to explore multiple alternatives is crucial to supporting creativity in design tasks~\cite{shneiderman2007creativity, hartmann2008design}.
For example, in the realm of interior design, designers typically develop a variety of versions to present to clients or stakeholders. 
To facilitate the exploration of multiple design variations, VRCopilot offers multiple empty workspaces or templates for users to work on (as seen in Fig. \ref{fig:authoring_interface_2}).
Users can easily switch between workspaces to work on different versions by navigating a list of miniatures in VR.
This creativity support is inspired by the concept of ``World in Miniature'' (WIM)~\cite{stoakley1995virtual} and recent work on version control in VR~\cite{zhang2023vrgit}.
To help users reuse partial layouts across different versions of the designs, VRCopilot also includes a copy \& paste feature, shown as additional buttons bound to the handheld menu.
This feature allow users to copy multiple objects and paste them either in the same workspace or other workspaces.

\subsection{Generative Model in VRCopilot}
We used ATISS~\cite{paschalidou2021atiss}, an open-source generative model for indoor scene synthesis using autoregressive transformers, as our generative model.
ATISS is trained using an open-source dataset of 3D models called 3D-FRONT \cite{fu20213d}, from which we also build our furniture catalog.
This model takes room dimensions parameters as prompts and generates reasonable furniture arrangements of the full-room layout.
It is also versatile for user inputs such as asking for a suggested placement of a given furniture item, or asking for a suggested furniture item for a given location.
We chose this model because it has been used as baseline models for work in the domain of indoor scene synthesis (e.g., \cite{feng2024layoutgpt, wei2023lego}).

We integrated ATISS in VRCopilot and can generate suggestions for full-room layouts on demand.
In VRCopilot, users can access the generative model via either voice commands or the catalog menu (as shown in Fig. \ref{fig:authoring_interface}).
Upon receiving the request, our system can fill the entire room by placing suggested furniture in the user's current workspace.
Users can delete the suggestions and also run the generative model repeatedly.
\lei{Our system supports multiple room sizes, shapes, and types (e.g., bedrooms and living rooms), and can be easily extended to support arbitrary room sizes and shapes (e.g., users can draw the room) and the backend generative model can adapt to these specifications automatically.}

\subsection{Multimodal Specification}
Existing immersive authoring tools enable users to directly manipulate virtual objects similar to how they would manipulate them in the physical world.
However, direct manipulation does not suffice for all needs during immersive authoring.
One clear weakness of direct manipulation is that it makes it difficult to identify or manipulate a potentially large sets of objects.
For example, there is a massive number of objects and styles in our furniture catalog (e.g., the catalog is based on 3D-FRONT that contains 7,302 furniture items with textures).
It is difficult for users to specify generating a chair with \textit{minimalist} style via direct manipulation.
On the other hand, the inherent ambiguity of natural language instructions makes it difficult to use pronominal reference to objects in the scene~\cite{cohen1992role}.
For example, it is hard for the system to understand which location the user is referring to when the user describes ``generate a chair here'' without additional contextual information.

Inspired by multimodal specifications in graphical interfaces such as ``Put-that-there''~\cite{bolt1980put}, VRCopilot allows users to use speech and simultaneous pointing to specify their creation needs, increasing the naturalness and utility of language description in the immersive environments. 
Our system can process users' natural language voice commands and categorize them into several possible intents:
\begin{itemize}
    \item \textit{Object Generation}: generating individual objects;
    \item \textit{Object Regeneration}: regenerate individual objects;
    \item \textit{Object Duplication}: duplicate selected objects;
    \item \textit{Scene Completion}: initiating a request for generating the full-room layout;
    \item \textit{Wireframe Generation}: initiating a request for generating the (see specifics in Section \ref{intermediate_representation});
    \item \textit{Wireframe Labelling}: assigning an object type to a wireframe (see specifics in Section \ref{intermediate_representation});
    \item \textit{Deletion}: delete selected objects.
\end{itemize}

\begin{figure*}[t!]
  \centering
  \includegraphics[width=\linewidth]{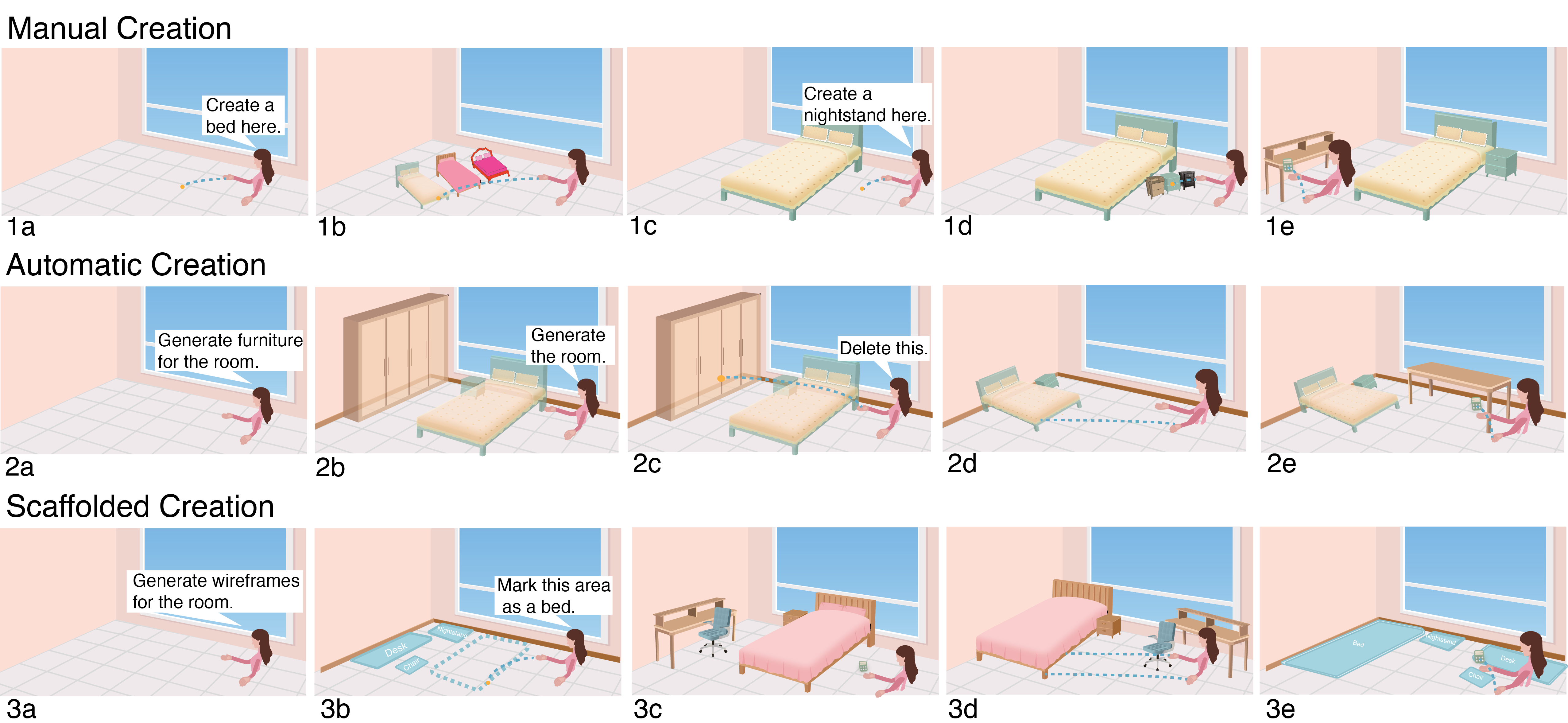}
  \vspace{-1.5pc}
  \caption{VRCopilot proposes three ways of human-AI co-creation in virtual immersive environments: \lei{manual} creation (1a-e), automatic creation (2a-e), and scaffolded creation (3a-e).}
  \Description[Three ways of human-AI co-creation in VRCopilot]{Figure 3 shows three ways of human-AI co-creation in virtual immersive environments: manual creation (1a-e), automatic creation (2a-e), and scaffolded creation (3a-e).}
  \vspace{-0.5pc}
  \label{fig:workflow}
\end{figure*}

Users can point to any location in the scene while verbally requesting that VRCopilot \lei{suggest} furniture to be placed at the designated point.
In their specifications, users can use pointing to specify the \textit{location} and voice to specify the \textit{object type} and its \textit{style} and \textit{material} (e.g., ``Generate a minimalist wooden chair here''), as seen in Fig. \ref{fig:workflow}.
Since its furniture catalog is built on 3D-FRONT, VRCopilot contains a limited number of 21 object types such as beds and chairs,  19 unique styles such as Modern and Japanese, and 15 unique materials such as Wood and Metal.
\lei{Our system’s language processing currently ignores out-of-range intents, such as indication of color or shape, due to the repository’s limitation of not supporting color or shape labels.
For example, when users indicated a *red* chair, the system would retrieve a chair of any color.
VRCopilot also does not include other natural language intents such as repositioning and object selection, as these operations can be more easily achieved via direct manipulation as observed in our pilot tests.}
Upon parsing the user's request, three suggested furniture items fitting the user’s provided criteria are visualized in front of the user (as seen in Fig. \ref{fig:workflow}), and the user can choose one of the three to become a part of the scene (as seen in Fig. \ref{fig:workflow}).

\subsection{Intermediate Representation}
\label{intermediate_representation}
One of the key challenges of human-AI co-creation is the lack of transparency, control, and user agency~\cite{amershi2019guidelines, buschek2021nine}.
To help users co-create with the generative model in a more transparent and controllable way, VRCopilot proposes the notion of \textit{wireframes} that is used as intermediate representations for the generated outcomes.
We took inspirations from low-fidelity prototyping that is commonly used in Human-Computer Interaction~\cite{rettig1994prototyping, buxton2010sketching, snyder2003paper}.
For example, prior work has explored using low-fidelity prototypes such as paper prototypes to quickly scaffold user interface design~\cite{snyder2003paper} or Play-Doh as intermediate representations to represent high-quality 3D models~\cite{nebeling2018protoar}. 
In VRCopilot, wireframes are designed as 2D representations of 3D layouts such as floor plans in interior design.
These representations can be hand-drawn by users together with speech specifying their types.
For instance, users can use the cursor of the raycast from the controller as the pen tip.
They can place the cursor on the floor and start drawing by pressing a button on the controller, while saying ``Mark this area as a bed.''
Upon the intent is recognized, the system will normalize the drawing into a rectangular plane with a text label of the object type (e.g. ``Bed'') attached to it. 
Users can further adjust the placement and dimension of the wireframe using direct manipulation, similar to manipulating furniture models.
Users can build up intermediate representation of the full-room layout design by creating multiple wireframes in the room.
Alternatively, users can ask the generative model to offer suggestions of wireframes by initiating a request similar to generating full-room layout.
The system can then generate the intermediate representation of the full-room layout and visualize all generated wireframes in the room.

In addition, VRCopilot allows users to iteratively refine the design with generative AI by enabling them to convert between intermediate representations and 3D layouts.
For example, users can use voice commands or button presses to turn their intermediate representations into actual furniture models.
The system can then interpret the labels and populates the scene with detailed furniture pieces corresponding to the \textit{object type}, \textit{size}, and \textit{orientation} as specified using each wireframe.
\lei{For objects that are not placed on the floor, such as ceiling lamps, users can draw wireframes on the floor similarly to how they create other objects. The system will then automatically set the \textit{y} attribute, representing the height, for these objects when populating the scene.}
Users can also switch back to intermediate representations from detailed furniture design, enabling an iterative design that leverages both lo-fi and hi-fi representations of the layout.

\subsection{Ways of Human-AI Co-creation in VR}
\label{three_ways}
With the above generative \lei{model} and interaction techniques, VRCopilot supports three ways of human-AI co-creation in VR.

\subsubsection{Manual Creation}
\lei{Manual} creation enables users to manually create a 3D layout design by creating each furniture item and its placement one after another.
Such creation method uses a bottom-up approach, where users start by creating specific furniture items either via the catalog menu or multimodal specification.
Once the central pieces are selected (e.g., beds, sofas), users consider how other components can be arranged within the room including placement of furniture, the flow of circulation, and how spaces will be utilized.

Figure \ref{fig:workflow} 1a-e showcases a typical workflow of how users can create layout designs using \lei{manual} creation: a) Users first use multimodal specification to ask the system to generate a bed. b) They can then pick from one of the suggestions as a central piece in the room. c) Users use multimodal specification to create a nightstand, and d) pick the one that best matches the style of the bed. e) Then they start using the catalog menu to create other furniture models such as desks to complete the layout design.

\subsubsection{Automatic Creation}
Automatic creation enables users to ask the generative model to generate full-room layouts.
After the suggested layout is given, users can modify the layout design based on their design goals.
This could include adjusting the placement of objects to avoid overlapping objects or to remove unwanted objects.

Figure \ref{fig:workflow} 2a-e demonstrates a typical workflow of how users can create layout designs using automatic creation: a) Users start out with no concrete design goals and ideas so decide to ask the system to generate the full-room layout for them. b) After the system processes the request and visualize the layout suggestion to the user, c) they can manipulate furniture models such as deleting the wardrobe to fit their own preferences. d) They decide to move the bed to be further away from the window. e) They also decide to add additional furniture models from the catalog menu that are missing from the generated layout such as additional desks. 

\subsubsection{Scaffolded Creation}
Scaffolded creation enables users to create intermediate representations, i.e., wireframes, to scaffold their designs.
Such a creation method uses a top-down approach where users begin with a broad, overarching vision of the floor plan by creating wireframes in the immersive environments.
They can draw their own wireframes and ask for generated wireframes.
They can also modify the placements and sizes of wireframes, and convert between wireframes and furniture layouts.

Figure \ref{fig:workflow} 3a-e a typical workflow of how users can iteratively create layout designs using scaffolded creation: a) Users first ask for generated wireframes from the system. b) Upon getting the results from the generative models, they can draw their own wireframes such as a bed and rearrange the wireframes. c) They can turn the wireframes into furniture layouts via a button press. d) They can then manipulate the furniture models to further fine-tune the design. e) Once they want to explore an alternative design, they can switch back to wireframes for generating another layout option.

\subsection{Implementation}
\begin{figure}[t!]
  \centering
  \includegraphics[width=\linewidth]{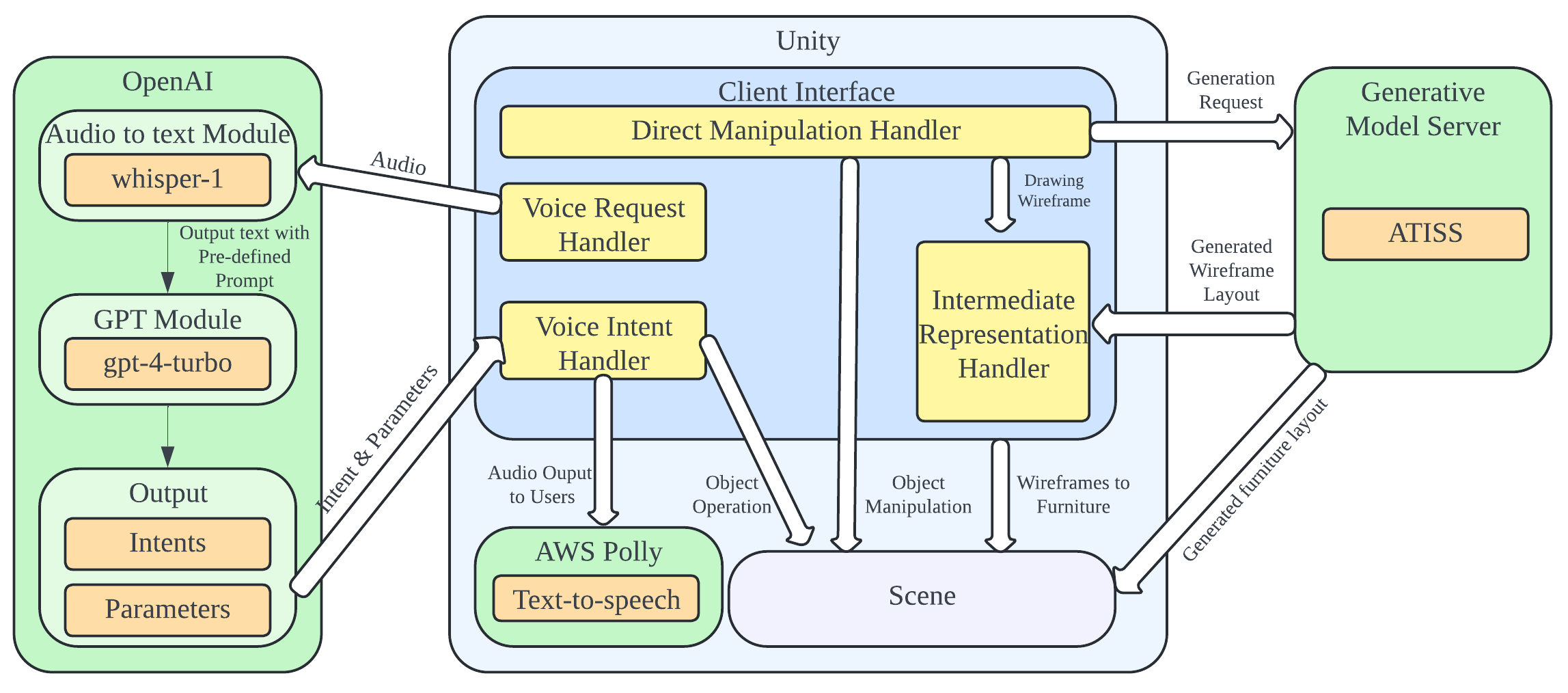}
  \vspace{-1pc}
  \caption{The system architecture of VRCopilot.}
  \Description[System architecture of VRCopilot]{Figure 4 shows the system architecture of VRCopilot.}
  \label{fig:implementation}
  \vspace{-.5pc}
\end{figure}

VRCopilot is developed using Unity (version 2021.3.20f1) and integrates plugins from Meta Oculus and the Microsoft Mixed Reality Toolkit (MRTK), enabling operation on Meta Quest and Rift VR headsets. The application incorporates advanced voice recognition and response capabilities through integration with the ChatGPT Audio Model (whisper-1) and Chat Model (gpt-4-turbo), with the latter hosted on a dedicated GPU server equipped with an Nvidia RTX 4090 graphics card. A comprehensive system architecture is depicted in Figure \ref{fig:implementation}.

\subsubsection{Integration with ChatGPT Models}
Interaction with ChatGPT models is facilitated through voice commands. The system captures user voice input via the microphone, converting the audio to an .mp3 format. This file is then translated into text by the ChatGPT SpeechToText model (whisper-1) through an HTTP request. The resulting text is processed by the ChatGPT Chat Model (gpt-4-turbo), which identifies the user's intent from the predefined categories and extracts relevant parameters such as furniture styles or categories. The responses, formatted as JSON, are parsed by the Unity client to execute the corresponding actions. While most actions are deterministic, actions requiring the generation of new items (e.g., ``generate a chair in a modern style'') involve a selection process from a set of items meeting the specified criteria.

\subsubsection{Communication with the Generative Model}
For tasks that involve the generation of new furniture, VRCopilot employs socket communication with a generative AI model, ATISS. Furniture attributes (unique ID, position, rotation, scale) are encoded in JSON and sent to the server. Upon completion, the server returns a JSON response with the furniture items that meet the established criteria, which the Unity client then processes and renders in the virtual environment.

\subsubsection{Multimodal Feedback Module}
To enhance user interaction, VRCopilot integrates a feedback loop through AWS Polly TextToSpeech model. After processing an intent, the system generates textual feedback corresponding to the user's request, which is then converted into speech. This multimodal feedback mechanism provides real-time auditory confirmation of actions taken within the virtual environment, enriching the user experience.

\section{User Study 1}
\label{evaluation_1}
To understand the effectiveness and challenges of co-creating with generative AI in immersive environments, we fist sought to compare immersive authoring with and without AI.
Prior research has provided some insights on how people collaborate with generative AI in creative domains (e.g. music~\cite{louie2020novice} and painting~\cite{chung2023promptpaint}).
We extend this line of work by understanding people's behaviors and attitudes when working with generative AI in virtual immersive environments.
We conducted a qualitative comparison study between two conditions: 1) immersive authoring using the conventional interfaces (e.g., via direct manipulation and menu selection), 2) immersive authoring with conventional interfaces and generative AI models.
We use this study as the first stop to eliciting the challenges that users perceived when co-creating with AI in VR.

\subsection{Participants}
We recruited 14 participants (10 women and 4 men, age 20–-28) from a university through public email lists. 
All participants had prior experience using VR devices and were compensated with \$30 USD Amazon gift cards for two hours of their time. 

\subsection{Procedure}
During the study, users were first given a tutorial of the system that covered individual features of the system including the control of direct manipulation and the usage of the generative model. 
The tutorial lasted about 30 minutes.
Then, participants were asked to design an empty apartment, consisting of two bedrooms and one living room, under two conditions: 1) with conventional immersive authoring interface, 2) with the conventional interface and the generative AI model.
\lei{The room sizes and types were pre-configured, in order to encourage participants to focus on the co-creation process.}
The order of the conditions was counterbalanced.
Each condition took about 15 minutes to complete.
In each condition, participants were asked to aim for finishing three versions of the apartment with at least three items in each room.
This instruction was not a strict requirement, but rather a means to encourage participants to design multiple variations of the apartment.
After both conditions were finished, we conducted a retrospective interview with participants.
Our study protocol was approved by our institution’s IRB.

\subsection{Analysis}
We transcribed and conducted a thematic analysis~\cite{braun2006using} of the interview data.
To assess the creation results from participants, we designed an evaluation that elicits emerging patterns of users' creation through an evaluation workshop.
One design expert, a full-time architect with 2 years of working experience, was invited to participate an evaluation workshop with one experimenter that took 90 minutes.
The evaluation workshop was held remotely where the experimenters screen shared to the expert.
The expert then went through the top down images of the creation results from all participants under each condition.
The order of showing the creation results is completely randomized and the expert was not informed of how the design were created under each condition.
Then the expert were asked to use an inductive approach to observe the top down images under each condition and use open-coding to elicit emerging patterns in each condition.

\subsection{Results and Insights}
Below are the insights gained from the qualitative user study and the expert evaluation:

\textit{Generative models provide less user agency.}
Agency refers to the awareness and control over one's action and their results~\cite{wilson2016nature}.
We found that participants reported feeling less agency over the creation results when co-creating with AI.
While the generative AI models could make meaningful layout suggestions that help users explore different ideas, the generated suggestions sometimes misaligned with users preferences in terms of the functionality and other considerations of the layout design.
For instance, P2 said ``I really have no idea what was going to come out when I did it [using the generative model], like I did not at all expect a bookcase in the middle of the living room, even if it would make sense for that room.''
P7 also commented on their agency when comparing creating with and without AI: ``When there's no AI intervention in the process it is just me thinking about what is the best circulations? What is the best looking furniture to be placed in the room? Those are my primary concern when I was doing it. So I will say I was the most in control when I was doing the first task [without generative AI models].''

\textit{Generative models are useful for sparking different ideas.}
One key dimension of creativity \lei{is} the ability to explore different ideas~\cite{cherry2014quantifying}.
Participants reported that the results from generative AI models can provide inspirations for the layout design that they did not come up with.
For example, P8 commented that ``It brings up new ideas I hadn't thought about before... also because when I first create the room, I pretty much put in what my favorite idea is for so when I create or generate a new room, it adds more inspiration than what I already had started off with.''

\begin{figure}[t!]
  \centering
  \subfloat[Creation Results without generative AI (P7).\label{fig:c1_1}]
  {\includegraphics[clip,width=0.48\linewidth]{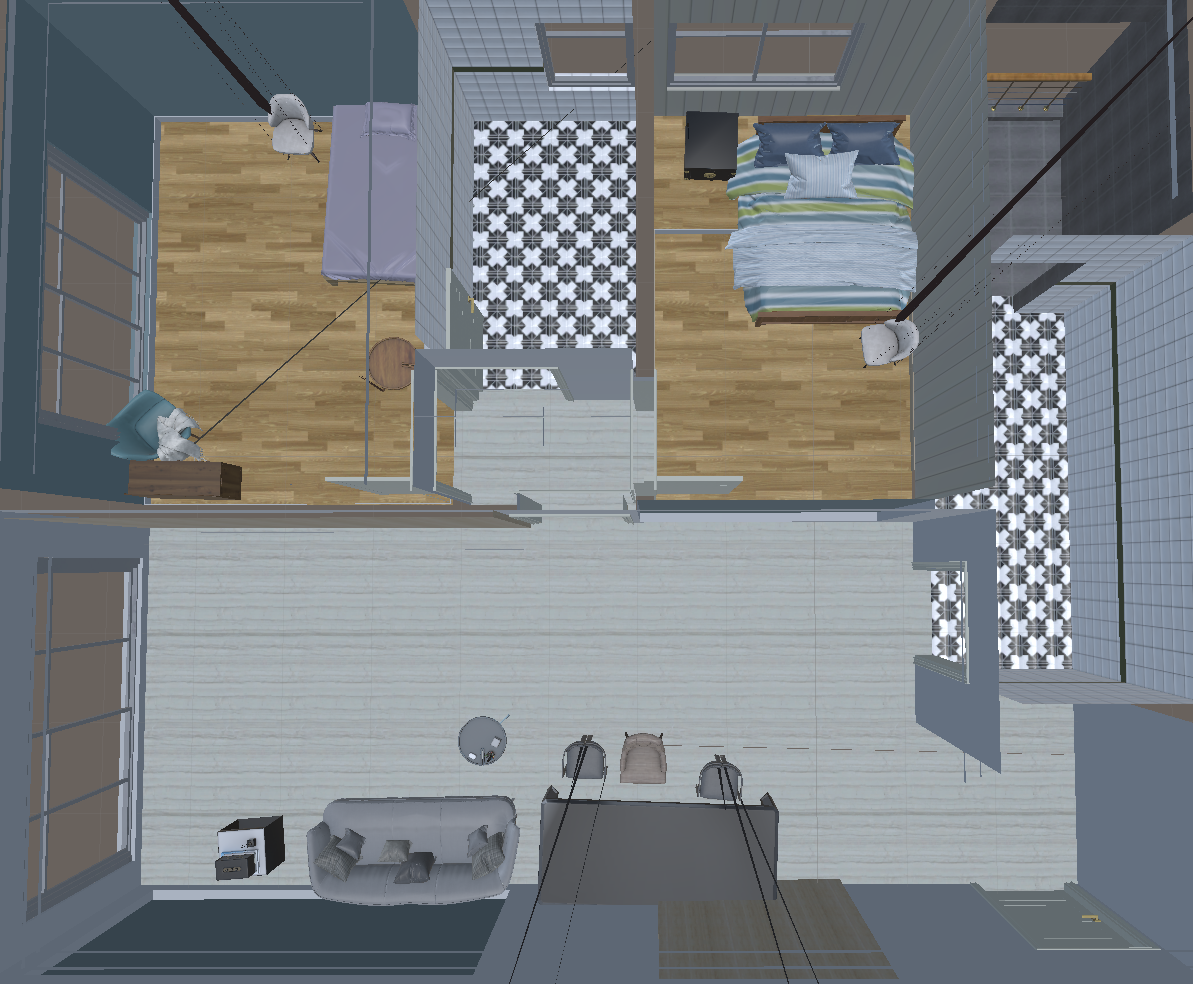}} 
  \hfill
  \subfloat[Creation Results without generative AI (P10).\label{fig:c1_2}]
  {\includegraphics[clip,width=0.48\linewidth]{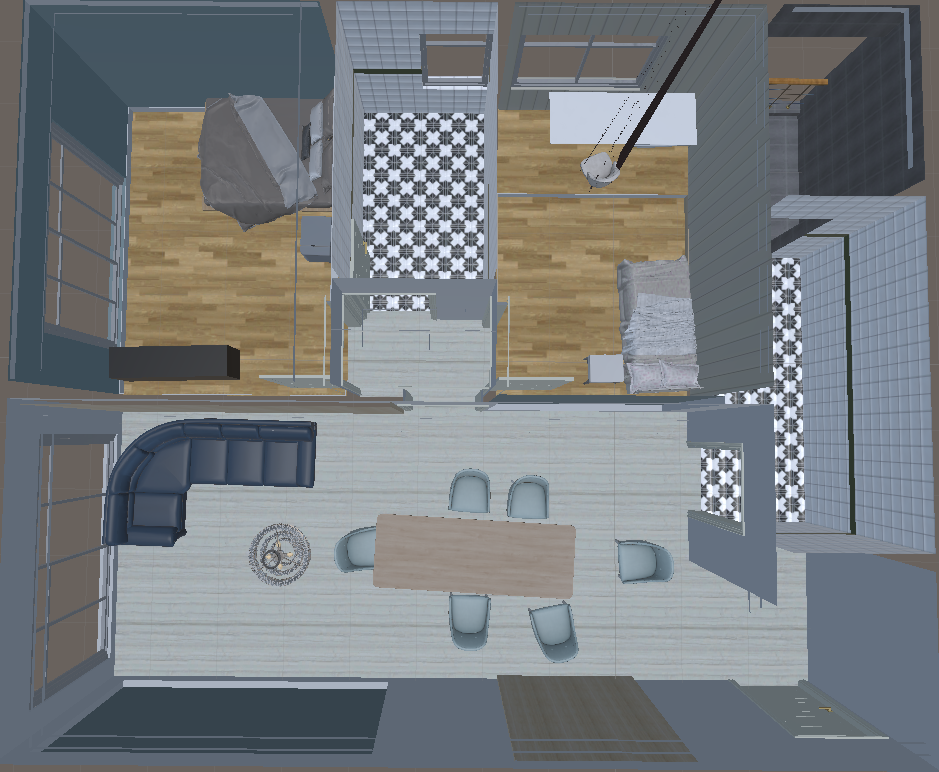}}
  \hspace{0pt}
  \subfloat[Creation Results with generative AI. Two bedrooms are created with generative AI while the living room is unfinished (P6).\label{fig:c2_1}]
  {\includegraphics[clip,width=0.48\linewidth]{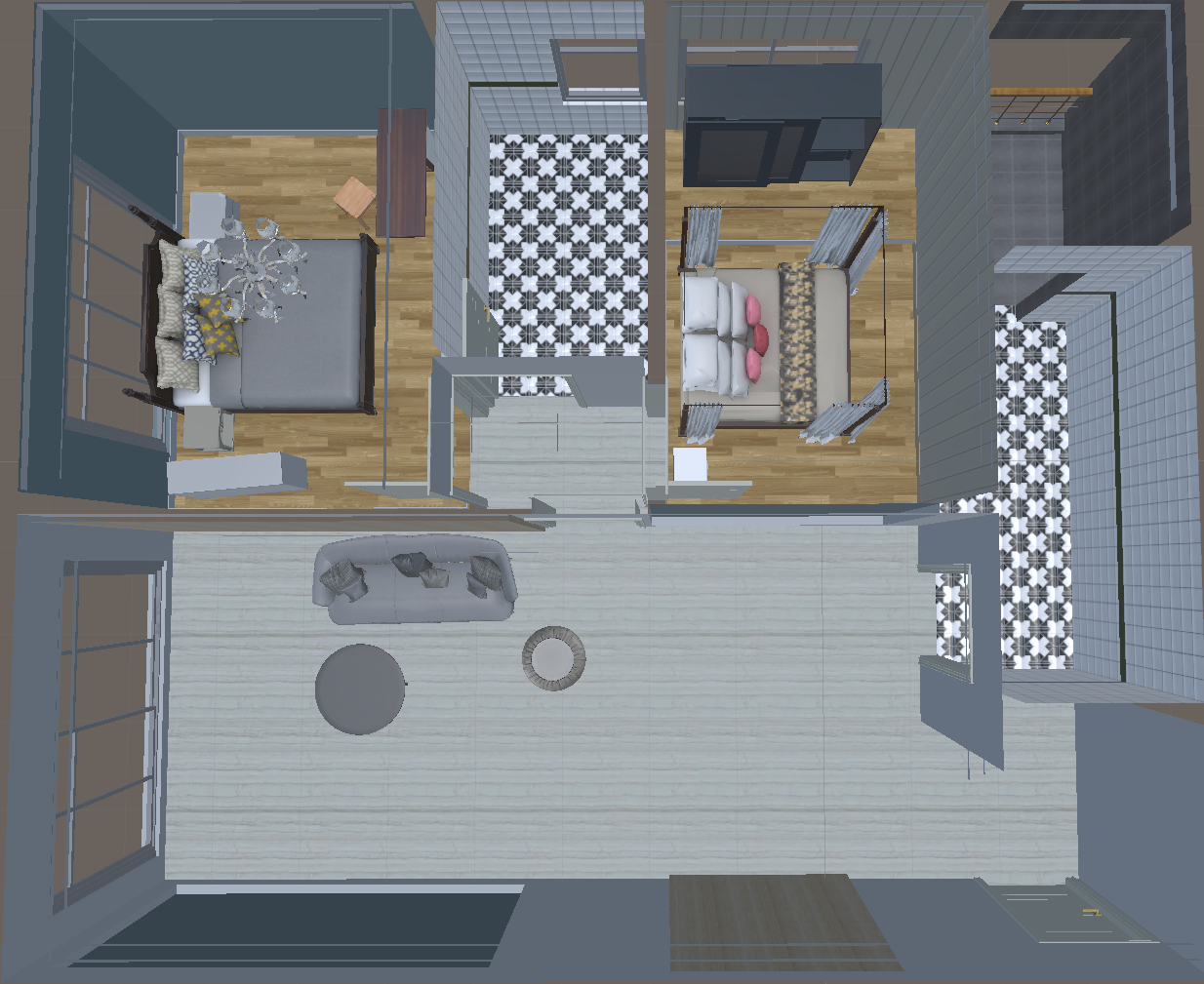}} 
  \hfill
  \subfloat[Creation Results with generative AI. All rooms are created with generative AI (P13).\label{fig:c2_2}]
  {\includegraphics[clip,width=0.48\linewidth]{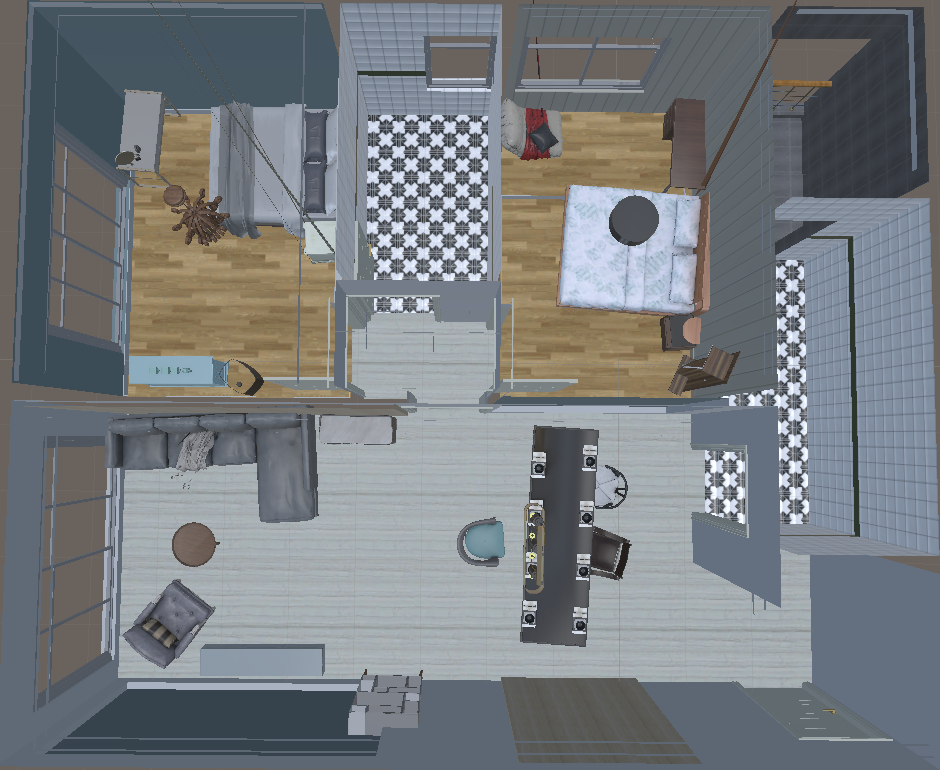}}
  \vspace{-.5pc}
  \caption{Exemplary top-down view comparison of participants' creation results with and without the assistance of generative AI in Study 1.}
  \Description[Top-down view comparison of participants' creation results in Study 1]{Figure 5 shows exemplary top-down view comparison of participants' creation results with and without the assistance of generative AI in Study 1.}
  \label{fig:creation_outcome}
  \vspace{-.5pc}
\end{figure}

\textit{Creating with generative models can lead to more diverse functionality and color palette.}
Functionality refers to the ability of a space or its components to serve a specific purpose or function effectively and efficiently.
We found that creation results with the help of AI encompass more diverse functionalities.
Specifically, the expert observed more diverse object types in each room that can support different activities.
For example, the bedrooms shown in both Fig. \ref{fig:c2_1} and Fig. \ref{fig:c2_2} include desks (for working), wardrobes (for clothes), and bookshelves (for storage).
Color palette refers to the selection of colors used in a design, including primary, secondary, and accent colors, which contribute to the overall mood and atmosphere of a space.
We found that creation results generated with AI generally have a richer color palette (seen in Fig. \ref{fig:creation_outcome}), which contributes to the expert commenting the creation ``more exciting.''

\textit{Creating with generative models can lead to poor considerations of circulation and daylighting.}
Circulation refers to the flow or movement of people within a space. It encompasses the pathways, routes, and patterns that individuals follow as they navigate and move through an interior environment.
We found that creation results generated with AI generally have a poorer circulation.
For example, one of the bedrooms shown in Fig. \ref{fig:c2_1} includes a nightstand that is blocking the doorway.
The dining table in the living/dining room in Fig. \ref{fig:c2_2} does not allow for much movement between the two sides due to its close placement to the walls.
This is because our underlying generative model (i.e., ATISS) that we utilize does not take doorway or room of movements into consideration when generating.
Daylighting in interior design is a design strategy that focuses on harnessing and optimizing natural daylight to illuminate interior spaces.
We found that creation results generated with AI generally have a poorer consideration of daylighting.
For example, both bedrooms shown in Fig. \ref{fig:c2_1} have furniture blocking the windows, making it difficult to harness daylight.
This is due to the underlying generative model (i.e., ATISS) that we utilize does not take window placement, size and shape into consideration.

Based on these findings around using generative models, our research team investigated further in the second round of study,  that was specifically focused on the mitigation of the issue of user agency and the comparison across different ways of human-AI co-creation (as described in Section \ref{three_ways}). 
We were also able to design tasks for the second study based on the patterns drawn from the expert evaluation session to further develop our ideas.
We describe the second user study in the following section.
\section{User Study 2}
We conducted a second user study to compare three conditions: 1) \lei{manual} creation using catalog menus and multimodal specification, 2) scaffolded creation using wireframes, and 3) automatic creation using generative AI.
We aimed to compare user perceived effort, creativity, and agency, and to elicit potential and challenges that users perceived when co-creating with AI in VR.

\subsection{Participants}
We recruited another 15 participants (5 women and 10 men, ages 19–-26) through university email lists. 
All participants had prior experience using VR devices and did not participate in the previous study (Section \ref{evaluation_1}).
We labeled the participants as P15-29 below.
Each participant was compensated with a \$30 USD Amazon gift card for two hours of their time.

\subsection{Procedure}
We designed a within-subject study where each participant experienced all three conditions during the study.
Balanced Latin-Square was used to determine the order of the conditions for each participant.
For the study setup, we used the Meta Quest 2 connected to a laptop that was running our system in Unity 3D game engine.
Each study session began with an introduction of the study and a tutorial of the system that lasted about 30 minutes. 
During the introduction, participants were introduced to the study and were informed of all the data that would be collected during the study. 
Participants were then given a tutorial of individual features of VRCopilot.
They were given an atomic task after learning each feature to familiarize themselves with the system.

After the tutorial, participants performed a design task under each condition, where they furnished an empty bedroom in VR.
In each condition, they were asked to come up with three design solutions for the same room within 15 minutes.
If more than three versions were created, they would be asked to turn in the three versions that they were most satisfied with.
The following design goals were given to participants for each condition:
\begin{itemize}
  \item There should be at least 4 furniture types in the bedroom.
  \item Make sure the top of the window is not blocked by wardrobes / shelves / bookcases.
  \item There should be enough space for users to navigate in the room.
  \item There should be a sofa to accommodate seating.
  \item Try to make the three versions different in both layouts and appearance.
\end{itemize}
The design goals were created based on design considerations drawn from the expert evaluation in the previous study (Section \ref{evaluation_1}) including functionality, day-lighting, navigation, seating, and diversity.
Participants were encouraged to design multiple variations of the room based on the design goals.
They were notified every five minutes during the task.
They were also free to ask for time remaining as well as clarification questions related to the task or the system.
However, experimenters were not allowed to give any instructions on how to design the room.
After finishing each task, participants were asked to fill out a questionnaire while we saved the resulted scenes and the screen recordings from that condition. 

After experiencing all conditions, we conducted a semi-structured interview with  participants to ask about user experience and their perceptions of each condition.
Each interview lasted about 20 minutes.
Our study protocol was approved by our institution's IRB.

\subsection{Measures and Analysis}
We evaluated the following metrics via post-task surveys.
Participants rated the items on a 7-point Likert scale (1=Strongly Disagree, to 7=Strongly Agree).
To answer the research questions, we measured the following aspects: 
(1) user perceived \textbf{effort} via the NASA-TLX~\cite{hart1988development} questionnaire; 
(2) user perceived \textbf{creativity} from the Creativity Support Index \cite{cherry2014quantifying}, with an emphasis on how well our system help users explore different ideas; 
and (3) user perceived \textbf{agency} adapted from prior work (e.g., Tapal et al.~\cite{tapal2017sense} and Lukoff et al.~\cite{lukoff2021design}).

We first performed a Friedman test, to analyze the non-parametric within-subject survey data.
For the potential post-hoc analyses, we conducted pairwise comparisons using Conover's test.
For qualitative results, we transcribed and conducted a thematic analysis~\cite{braun2006using} of the interview data.

\subsection{Quantitative Results}
\begin{figure}[t!]
  \centering
  \includegraphics[width=\linewidth]{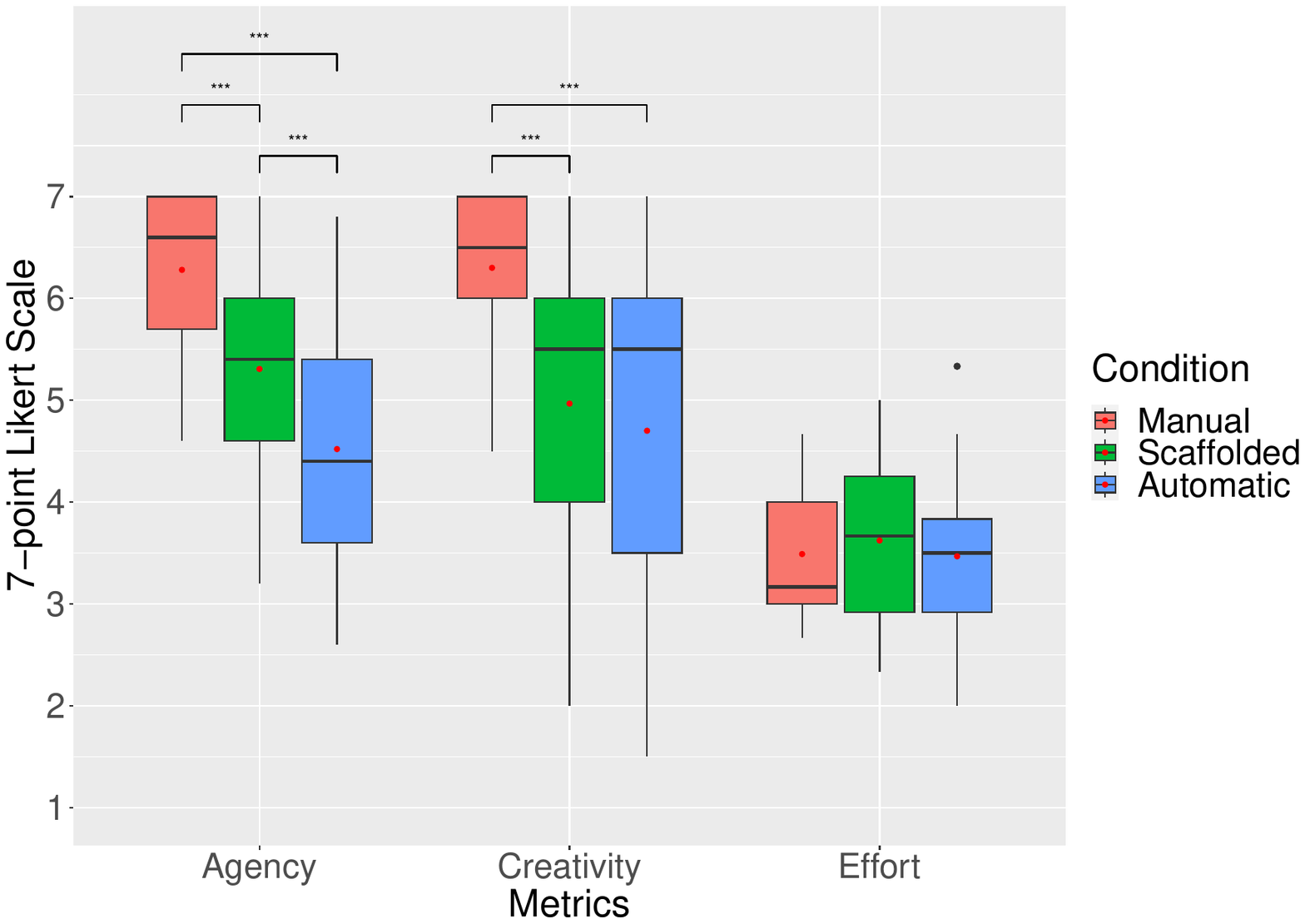}
  \vspace{-0.5em}
  \caption{Results from post-task survey comparing three conditions in Study 2.}
  \Description[Quantitative results from post-task survey in Study 2]{Figure 6 shows quantitative results from post-task survey comparing three conditions in Study 2.}
  \label{fig:rplot}
\end{figure}

The results of the post-task questionnaire and the Conover’s test are aggregated and shown in summarized in Figure \ref{fig:rplot}.

A Friedman test was conducted to evaluate differences in participants' perceived agency across three conditions. 
The analysis revealed a statistically significant difference in the sense of agency across the three conditions (\( \chi^2(2) = 20.11, df = 2, p < .001 \)).
Post-hoc analyses with Conover's pairwise comparisons were performed with a Bonferroni correction.
We found that participants' perceived control was significantly higher in the \lei{manual} creation condition (\(p < .001 \) compared to the scaffolded creation condition and \(p < .001 \) compared to the automatic creation condition).
We also found that participants' perceived control was significantly higher in the scaffolded condition compared to the automatic condition (\(p < .001 \)).
These results suggest that the design of wireframes was effective in increasing users' sense of control compared to fully automatic generation from AI.

In regards to users' perception of creativity, we found a significant effect of ways of creation on the sense of creativity (\( \chi^2(2) = 17.633, df = 2, p < .001 \)).
Further post-hoc analysis revealed that users felt significantly higher sense of creativity in the \lei{manual} condition compared to both the scaffolded condition (\(p < .001 \)) and the automatic condition (\(p < .001 \)), with no significant difference found between the latter two.
These findings show that participants felt they were the most creative when they were working in the \lei{manual} creation condition, but similarly creative in the scaffolded and automatic creation condition.

For users' perceived effort, we did not find a significant effect of ways of creation on users' perceived effort (\( \chi^2(2) = 0.915, df = 2, p = 0.63 \)).
This indicates that users felt similar levels of effort across three conditions.

\subsection{Qualitative Results}
All participants were able to finish three design variants of the room by the end of the task.
To further investigate the reasons behind users' perceptions in subjects such as agency and creativity, we analyzed our interview data and solicited users' qualitative feedback.
\lei{Overall, our results suggest that users' sense of agency can be enhanced by offering greater control during the human-AI co-creation process, such as control over object types and sizes through wireframes or object styles through multimodal specification.
In addition, providing multiple suggestions via multimodal specification can increase users' creativity.
We center our findings below around the topics of user agency and creativity in the contexts of human-AI co-creation, interior design, and immersive authoring.}

\subsubsection{Offering greater control could enhance the sense of agency and ownership.}
Participants reported having higher agency over the created content in scaffolded creation compared to in automatic creation.
Specifically, participants felt that they have control over aspects such as furniture size and placement compared to automatic creation.
\begin{quote}
    ``I felt like I had the most control with the wireframe because in addition to what types of furniture I could also decide what size and how it's positioned. Whereas the others, I think, particularly lost out on the sizing component. Because there were a few times, after I tried the wireframe, that I did try to resize furniture, but then realized that that wouldn't work [in the other conditions].'' -P23
\end{quote}

\lei{In addition, participants reported having higher agency in manual creation since they have additional control over the furniture styles.}
\begin{quote}
    \lei{``For example, I can pick, only leather chairs, leather sofas, and then have a bed that matches that style... you just got more control over the style itself, rather than just the layout" -P21}
\end{quote}

We also found that users felt the least agency in the automatic creation since the generated furniture is already fleshed out and decreases their willingness to control or manipulate things, which further decreased the sense of ownership in the created content.
\begin{quote}
    ``Because it feels like that's already there. So it looks like it already looks pretty good. So I wouldn't want to move it too much, and definitely I have less control with it. Because the furniture and everything were chosen by AI, I feel like it doesn't feel fully like \textit{I} designed it.'' -P28
\end{quote}

\subsubsection{\lei{Manual} creation can spark creativity via multiple suggested options.}
We found that participants felt the most creative in the \lei{manual} creation, mostly because the multiple suggestions offered via multimodal specification can give users inspirations.
Having the options from the system could inspire participants to keep building the room centering around the piece that they chose from the options.

\begin{quote}
    ``When I saw the bed [from the three suggestions], and it's like bright green, yellow, I was like, `maybe I can make this the theme of this room.' And I was trying to go with this style when I was choosing the other furniture. Then when I saw a bunk bed, I was like, `maybe this could be a bunk bed for two children,' and I'm styling the room in that way.... I think the [\lei{Manual} Creation] condition facilitates creativity a bit more just because you can choose between the three options.'' -P24
\end{quote}

\subsubsection{Users tend to follow the designs generated in scaffolded or automatic creation, leading to a reduced sense of creativity.}
While the scaffolded creation and the automatic creation also suggest furniture to the user, participants reported less creativity mostly because they tended to follow the layout that the generative model suggests to them.
We found that users tended to feel fixation when the system generated the full-room layout compared to the system suggested individual furniture items.

\begin{quote}
    \lei{``I think having everything laid out for you already, it decreases your creativity. Because you'll have that bias towards the way that it just puts everything. So it's like the bed's here, I might just keep it there." -P21}
\end{quote}

\subsubsection{Scaffolded creation enables high-level and unbiased design thinking.}
Participants mentioned that scaffolded creation, specifically the design of wireframe, allowed them to focus on the functionality over the styles and enabled them to think ``in the layout sense'' (P24).

\begin{quote}
    ``I feel like, by creating all those wireframes, I'm actually doing the job of an interior designer, because I'm not the one who's purchasing the actual furniture for the household. I'm just designing how to maximize the utility of the whole space for this household.'' -P20
\end{quote}

\begin{quote}
``I think just where things are and how you move around the room. I think that's very important...if the room is cramped or awkward, it's not going to be as good even if it looks really nice. So wireframe, I think, is very good for that just to see it completely unbiased. Because if I just build using the voice or the menu, I can already see things. Like if it looks good, but it's not really functional, I might be biased just because it looks good, and just go with something that doesn't really work. But wireframe kind of takes away from that. And it really lets me focus on the function. And just making sure that everything flows together nicely.'' -P19
\end{quote}

\subsubsection{\lei{The design of wireframes in scaffolded creation enables easy manipulation in VR}}
\lei{We found that the design of wireframes in scaffolded creation made it easier for users to navigate the layout and manipulate distant objects due to the reduced occlusion of 2D planes, compared to handling a full layout with 3D furniture.}

\begin{quote}
\lei{``I think it [wireframe] is useful in, getting through the layout, because with all the objects already in the environment, it's been hard to see around and if there's something behind the big cabinet, you can't reach it. But with wireframes, you can see everything at once.''  -P18}
\end{quote}

\subsubsection{\lei{Expectation mismatches with system suggestions reduce user control}}
\lei{We found that users sometimes felt that the system's suggestions, either suggested via multimodal specification or generative AI models, did not match their design expectations, and thus reduced their sense of control.
For example, participants were not able to specify the color or the relative size (e.g. big or small) of the objects either through multimodal specification or wireframes.
This kind of mismatch is often due to the lack of understanding of the capabilities of the underlying AI models.}

\begin{quote}
    \lei{``When I was trying to create a side table to place next to a sofa as a coffee table, either it was not picking up or it was going for more desk or larger-size tables. Even though I switched back and forth between saying small table and side table, it still took a while before it generated something I was happy with." -P23}
\end{quote}
\section{Discussion}
In our first study, we found that generative models are helpful for idea inspirations.
Through the followup expert elicitation study, we found that when co-creating with generative AI models, users can create 3D layouts with more diverse functionality and color palette, but with poorer consideration of circulation and daylighting.
Furthermore, we found that generative models could result in lower user agency when it comes to human-AI co-creation in immersive environments.
However, this could be mitigated via the design of wireframes as found in our second study.

Our second study demonstrated that among the three ways of human-AI co-creation, \lei{manual} creation offers users the most sense of agency and creativity.
By visualizing multiple \lei{furniture} suggestions, \lei{manual} creation can offer design inspirations.
Scaffolded creation offers users higher agency compared to automatic creation.
This is because users have additional control over aspects such as furniture size and placement via scaffolded creation.
Users also found that scaffolded creation can enable un-biased, higher-level of thinking when designing layouts.
In automatic creation, users tended to follow what the system suggested to them and not to make changes, leading to the least sense of creativity and agency among the three conditions.

\subsection{Design Implications}
Through the lens of creativity and agency, we highlight the opportunities and challenges of human-AI co-creation in immersive virtual environments, and discuss design recommendations drawn from our results.

\subsubsection{Offering results of generative AI via intermediate representations}
Our design of wireframes offers higher user controllability when working with generative models.
Specifically, in the task of creating 3D layouts, users are granted more control over the size and placement of object and think they can view the design in an unbiased way.
\lei{Besides, the design of wireframes offers unique affordances in VR by making it easier for users to navigate layouts and manipulate distant objects due to less occlusions compared to handling a fully populated 3D scene.}
This aligns with prior work that utilizes low fidelity representation when working the generative designs (e.g. ~\cite{kazi2017dreamsketch}).
Similarly, there has been also a long-standing body of work in Sketch Based Interfaces for Modeling that utilizes both the coarseness and the expressiveness of sketches to guide the detailed generation of 3D models~\cite{olsen2009sketch}.
This demonstrates the benefits of designing low fidelity representations that can prompt more controllable and sophisticated generated content.
\lei{We therefore encourage future researchers and designers to consider using more advanced intermediate representations of the generated outcome beyond 2D planes on the floor.
These representations should capture richer properties of 3D content, such as color and shape, in the immersive environment while still allowing users to easily manipulate the objects and navigate the scene.
The note of intermediate representations could even go beyond immersive environments.
The concept of intermediate representations can extend beyond immersive environments. For instance, rather than generating lengthy text, Large Language Models could produce an outline as an intermediate representation, allowing users to make adjustments before finalizing the text. 
Similarly, other generative models could use intermediate representations like image skeletons for pictures or key frames for videos.}

\subsubsection{Offering multiple generated suggestions for inspirations}
Our study shows that participants felt more creative \lei{and} more easily inspired when they can choose from multiple generated suggestions.
Users tend to get inspirations from suggestions when they don't have a concrete idea in mind or when they don't want to spend too much time on browsing the catalog menu.
Contrarily, when given one suggestion in automatic creation, users tended to follow what the system generates, leading to fixation of thinking.
Thus, future researchers and practitioners might consider offering user the ability to choose from multiple generated suggestions, in order to enhance users' sense of creativity.
For example, generative AI systems can offer parallel comparison by visualizing potentially diverse generated results for users.

\subsubsection{Addressing expectation mismatch between users and generative AI}
A common challenge across all conditions, based on the study, is the expectation mismatch when unexpected output was generated by AI. 
Through the expert evaluation, we found that although by co-creating generative AI models users can create 3D layouts that are diverse in aspects such as functionality and color palette, users generally have preferences of the layout design that fall outside of the capabilities of generative AI models.
\lei{For example, in study 1, layouts co-created with AI showed poorer consideration of circulation and daylighting because the underlying generative AI model was not trained with those criteria in mind. 
Additionally, users lacked a sufficient understanding of the system's capabilities. 
This highlights the need for more transparent communication between users and generative AI regarding the system's capabilities and limitations.}
This aligns with the Explainable AI (XAI) research (e.g., ~\cite{liao2023designerly, dhanorkar2021needs, guidotti2018survey}), where researchers aim to provide more transparent explanations of decision-making process of the AI model, with an emphasis on text or images.
However, there has been little explorations in the visualization and interaction techniques for making AI models more understandable in the immersive environments.
Therefore, future researchers and practitioners should consider designing human-AI systems that can visualize how the generative AI model perceives and completes the user's design. 

\subsection{Limitations}
Our paper explored ways of human-AI co-creation in virtual immersive environments and showed empirical results on the comparison among various ways of creation.
However, our work has several limitations.
First, both of our studies took place in a lab setting with the participants engaged with the system in a short amount of time.
The way that participants design 3D layout with a time constraint in the lab setting could be different from how they would design without time constraint outside the lab. 
Our studies also had a relatively small sample size, which could reduce the validity of our quantitative results.
\lei{Second, our system was specifically tailored for interior design tasks and had several technical limitations. 
For instance, as mentioned in Section 3, users could only draw wireframes on the floor, and for objects not placed on the floor (e.g., ceiling lamps), the system automatically set their heights when converting to furniture.
The underlying AI models of VRCopilot also had limitations, such as misidentifying voice intents or not supporting color or size in multimodal specifications. 
Participants occasionally had to retry their intents or regenerate in a few cases when the system misidentified voice commands.
Future work should seek to provide clearer system status and offer alternatives for misidentified or unsupported intents, as well as further extend the model’s capabilities and supported attributes.
Lastly, the generalizability of our findings to domains or contexts other than interior design necessitates further investigation. 
Our paper provides insights into user creativity, agency, and strategies in human-AI co-creation in general.
Some findings, however, are more specific to interior design or the immersive virtual environments.
Future research should evaluate the adaptability and utility of the system across diverse application domains to determine its broader applicability.}

\section{Conclusion}
In this paper, we presented a mixed-initiative system named VRCopilot that integrates pre-trained generative models into immersive authoring workflows.
We introduce three ways of human-AI co-creation in the immersive virtual environment including \lei{Manual} Creation, Automatic Creation, and Scaffolded Creation.
We conducted two rounds of comparative studies that evaluates the potential and challenges of co-creating with generative AI in VR and user perceived creativity, effort, and agency.
Our first study revealed that generative AI could offer design inspirations to users but decrease their sense of agency.
Our second study suggested that when users use the wireframes in Scaffolded Creation, they felt higher sense of agency compared to Automatic Creation.
\lei{Manual} Creation offers users the most creativity and agency.
We provide insights on the opportunities and challenges around human-AI co-creation in the immersive environments and make recommendations for future research and design.

\begin{acks}
We would like to thank Bella Palumbi for her help in system implementation. We would also like to thank our participants for their time, and the reviewers for their valuable feedback and suggestions.
\end{acks}

\bibliographystyle{ACM-Reference-Format}
\bibliography{references}

\appendix

\end{document}